\documentclass[]{aastex631}

\usepackage{amsmath}
\usepackage{graphicx}

\usepackage{comment}

\newcommand{\VEC}[1]{{\bf {#1}}}

\begin{document}


%
\title{Accumulation of elastic strain toward crustal fracture in magnetized neutron stars}
%
%
%
\author[0000-0002-4402-3568]{Yasufumi Kojima}

\affiliation{
Department of Physics, Hiroshima University\\
Higashi-Hiroshima,739-8526, Japan}

\begin{abstract}
This study investigates elastic deformation driven by the Hall drift in a magnetized neutron-star crust.
Although the dynamic equilibrium initially holds without elastic displacement, the magnetic-field evolution changes the Lorentz force over a secular timescale, which inevitably causes the elastic deformation to settle in a new force balance.
Accordingly, elastic energy is accumulated, and the crust is eventually fractured beyond a particular threshold.
We assume that the magnetic field is axially symmetric, and we explicitly calculate the breakup time, maximum elastic energy stored in the crust, and spatial shear-stress distribution.
For the barotropic equilibrium of a poloidal dipole field expelled from the interior core without a toroidal field, 
the breakup time corresponds to a few years 
    for the magnetars with a magnetic field strength of $\sim 10^{15}$~G;
however, it exceeds 1 Myr for normal radio pulsars.
The elastic energy stored in the crust before the fracture ranges from $10^{41}$ to $10^{45}$ erg, 
depending on the spatial-energy distribution.
Generally, a large amount of energy is deposited in a deep crust.
The energy released at fracture is typically $\sim 10^{41}$ erg when
the rearrangement of elastic displacements occurs only in the fragile shallow crust.
The amount of energy is comparable to the outburst energy on the magnetars.
\end{abstract}

%
%

%
\section{Introduction}

Neutron-star crust is considered a key aspect for understanding several astrophysical phenomena.
The solid layer near the stellar surface can support non-spherical deformations, called mountains,
with a height of less than 1 cm.
Such asymmetries on a spinning star cause the continuous emission of gravitational waves.
Therefore, calculation of the maximum size of these mountains is key to the detection of gravitational waves and has been discussed in several theoretical studies 
\citep[][]{2000MNRAS.319..902U,2004MNRAS.351..569P,
2006MNRAS.373.1423H,2021MNRAS.500.5570G}.
Gravitational-wave observation provides valuable information 
\citep[][for recent upper limit]
{2021ApJ...921...80A,2021ApJ...913L..27A,2022arXiv220404523T};
thus, by continuously improving the sensitivity of the LIGO-Virgo-KAGRA detectors,
 the physics relevant to the phenomenon may be explored.

Pulsar glitches are sudden spin-up events that are observed in radio pulsars
\citep[][for glitch catalogue]{2011MNRAS.414.1679E,2022MNRAS.510.4049B}.
Similar spin-up and peculiar spin-down events are observed in anomalous X-ray pulsars
\citep[][]{2008ApJ...673.1044D,2017ARA&A..55..261K}.
A sudden spin-up in a radio pulsar is produced by the transfer of angular momentum from
the superfluid components of the core to the normal crust
\citep[][]{1975Natur.256...25A,1984ApJ...276..325A}.
Crust quakes were also discussed in other models 
\cite[][for recent studies]
{2000ApJ...543..987F,2020MNRAS.491.1064G,2021A&A...654A..47R}.
The elastic deformation is caused by a decrease in centrifugal force, owing to a secular spin-down,
and the crust eventually fractures when the strain exceeds a critical threshold.
However, this simple model does not explain the observation; the loading of the solid crust between glitches is too insignificant to trigger a quake.
%

Giant flares in magnetars are rare, albeit highly energetic. They are typically
$\sim 10^{44} -10^{46}$ erg released within a second
\citep[][for a review]
{2015RPPh...78k6901T,2017ARA&A..55..261K,2021ASSL..461...97E}.
Quasi-periodic oscillations (QPOs) with some discrete frequencies in the range of 20 Hz -- 2 kHz 
were observed in the tails of these flares.
Per the order-of-magnitude estimate, these frequencies correspond to the torsional shear or 
 {Alfv{\'e}n} modes with magnetic field strength $\sim 10^{15}$ G.
Outbursts, which are less energetic, are also observed in magnetars.
These activities are considered to be powered by internal strong magnetic fields of
$\sim 10^{15}$ G.
The crustal fracture of the magnetar is proposed as a model for fast radio bursts (FRBs)
\citep[][]{2019MNRAS.488.5887S,2020ApJ...903L..38W}, 
and it may be supported by QPOs \citep[][]{2022ApJ...931...56L}
in the radio burst from SGR J1935+2154 in the galaxy
\citep[][]{2020ApJ...898L..29M, 2020Natur.587...54C, 2020Natur.587...59B}.
Most FRBs are located at a cosmological distance, and further observation sheds light on whether FRBs originate from magnetars or a subclass.
A recent observation of the magnetar SGR 1830-0645 revealed pulse-peak migration  
during the first 37 days of outburst decay \citep[][]{2022ApJ...924L..27Y}.
This provides important information concerning the crust motion coupled with the
exterior magnetosphere.

Most theoretical studies have been focused on the crustal-deformation limit.
Elastic stresses gradually accumulate until a particular threshold.
Beyond this threshold, the elastic behavior of the lattice abruptly ceases, and
 the transition is exhibited as a star-quake or burst.
An evolutionary calculation of the deformation
is necessary to understand the related astrophysical phenomena.
%

In this study, we consider the crust in a magnetized neutron star.
The static magneto-elastic force balance was studied for various magnetic-field configurations
\cite[][]{2021MNRAS.506.3936K,2022MNRAS.511..480K}.
 A variety of magneto-elastic equilibria was demonstrated, and is considerably different from the
 barotropic equilibrium without a solid crust.
Herein, we explore the accumulation of shear stress induced by the Hall evolution, which is an important process 
in the strong field-strength regime.
Suppose that the
 magneto-hydro-dynamical (MHD) equilibrium in the crust holds at a particular time without the elastic force. 
The equilibrium is not that for electrons
\citep[][]{2013MNRAS.434.2480G,2014MNRAS.438.1618G}; thus,
 the magnetic field tends to the Hall equilibrium in a secular timescale.
According to the magnetic-field evolution, 
the Lorentz force also changes.
The deviation is assumed to be balanced with the elastic force.
Thus, the shear stress in the crust gradually accumulates 
and reaches a critical limit. 
We cannot follow the post-failure evolution because some uncertainties are involved in 
the discontinuous transition.
Therefore, our study provides the recurrent time and magnitude of the bursts.

The models and equations used in the study are discussed in Section 2. 
For MHD equilibrium in a barotropic star,
 the evolution of the magnetic field is driven by a spatial gradient of electron density.
In Section 3, the critical configuration at the elastic limit is evaluated, and
the accumulating elastic energy is calculated.
In Section 4, we also consider non-barotropic effects using simple models.
The non-barotropicity results in another driving process of the magnetic-field evolution, and
consequently elastic deformation.
The numerical results of these models are given.
Finally, our conclusions are presented in Section 5.

\section{Formulation and Model}
  \subsection{Magnetic Equilibrium}

We consider the dynamical force balance between pressure, gravity, and the Lorentz force. 
The MHD equilibrium is described as follows:
\begin{equation}
-\frac{1}{\rho}{\VEC {\nabla}}P-{\VEC {\nabla}}\Phi_{\rm g}
+\frac{1}{c\rho} {\VEC {j}}\times {\VEC {B}}=0,
  \label{Forcebalance.eqn}
\end{equation}
where $\Phi_{\rm g}$ is the gravitational potential including the centrifugal terms. 
The third term has a magnitude $\sim 10^{-7}(B/10^{14}{\rm G})^2$ times
smaller than those of the first and second terms.
Deviation owing to the Lorentz force is small enough to 
be treated as a perturbation on a background equilibrium.

We limit our consideration to an axially symmetric configuration for the magnetic-field configuration.
The poloidal and toroidal components of the magnetic field are expressed by two functions $\Psi$ and $S$, respectively, as follows:
\begin{equation}
{\VEC{B}}={\VEC{\nabla}}\times \left(\frac{\Psi}{\varpi}{\VEC{e}}_{\phi}\right)
+\frac{S}{\varpi}{\VEC{e}}_{\phi},
 \label{eqnDefBB}
\end{equation}
where $\varpi=r\sin\theta$ is the cylindrical radius, and
$\VEC{e}_{\phi}$ is the azimuthal unit-vector in $(r,\theta, \phi)$ coordinates.
When the equilibrium is barotropic,
i.e., the constant surfaces of $\rho$ and $P$ are parallel, 
 the azimuthal current $j_{\phi}$ is described in the form 
\begin{equation}
\frac{4\pi j_{\phi}}{c} = \rho \varpi\frac{dK}{d\Psi}
+\frac{S}{\varpi}\frac{dS}{d\Psi},
 \label{MHDeqil.eqn}
\end{equation}
where the current function $S$ should be a function of $\Psi$, and $K$ is another function of $\Psi$.
For the axially symmetric barotropic equilibrium,
acceleration due to the Lorentz force, which is abbreviated as 
${\VEC {a}} \equiv (c\rho)^{-1}{\VEC {j}}\times {\VEC { B}}$, is given by
\begin{equation}
 {\VEC {a}}=\frac{1}{4\pi}{\VEC {\nabla}}K.
 \label{EMForcebaro.eqn}
\end{equation}
The force balance (\ref{Forcebalance.eqn}) is
described by gradient terms of scalar functions.

The magnetic function $\Psi$
is obtained by solving the Amp{\'e}re--Biot--Savart law with the source term (Equation~(\ref{MHDeqil.eqn})) after the functional forms $S(\Psi)$ and $K(\Psi)$ are specified. 
For simplicity, we assume that
$K$ is a linear function of $\Psi$,
$K=K_{0} \Psi$, and $S=0$ in  Equation~(\ref{MHDeqil.eqn}).
The poloidal magnetic field is purely dipolar and is given by $\Psi=g(r)\sin^2\theta$.
The radial function $g$ is solved with appropriate boundary conditions:
in a vacuum at the surface $R$, and $g=0$ at the core-crust interface $r_{c}$.
The latter refers to the magnetic field expelled from the core.
For the case where the field penetrates into the core,
$g$ smoothly connects to the interior solution at $r_c$.
We normalize the radial function
by the dipole field strength $B_{0}$ at the surface, $g(R)=B_{0}R^2/2$.
The magnetic geometry discussed above is a simple initial model to examine the magnetic field evolution.
However, purely poloidal magnetic field configurations are unstable according to an energy principle
\citep[][]{1973MNRAS.161..365T,1973MNRAS.163...77M,1973MNRAS.162..339W} and
numerical MHD simulation 
\citep[][]{2006A&A...450.1077B,
2009MNRAS.397..763B,2011MNRAS.412.1394L,2011MNRAS.412.1730L,
    2015MNRAS.447.1213M}.
Dynamical simulations revealed that the final state after a few 
 {Alfv{\'e}n}-wave crossing times is a twisted-torus configuration, in which the poloidal and toroidal components of comparable field strengths are tangled.
Moreover, recent three-dimensional simulation shows very asymmetric equilibrium
\citep[][]{2022MNRAS.511..732B}.
These studies are concerned with the configuration in an entire star.
The information relevant to our study corresponds to the magnetic field in a thin outer layer; therefore, the present understanding is quite incomplete.
For example, 
the ratio of toroidal to poloidal components decreases near the surface,
because the toroidal field should vanish outside the exterior. 
However, the ratio in the crust located near the surface is uncertain, almost zero or in the order of unity, although both components are comparable in magnitude inside the star.
Initially, a simple magnetic field configuration is used in this paper, but it is necessary to improve the configuration.
We discuss the non-barotropic equilibrium of the magnetic field.
From Equation~(\ref{Forcebalance.eqn}), 
the acceleration owing to the Lorentz force satisfies 
\begin{equation}
      {\VEC {\nabla}}\times {\VEC {a}}=
       {\VEC {\nabla}} \times 
          \left(\frac{1}{\rho} {\VEC {\nabla}}P \right) \ne 0.
          \label{nonbarac.eqn}
\end{equation}
The acceleration ${\VEC {a}}$ is no longer described by the gradient of a scalar, but
 may be generalized as
\begin{equation}
 {\VEC {a}}=\frac{1}{4\pi}{\VEC {\nabla}}K+\alpha {\VEC {\nabla}}\beta,
 \label{EMForceNB.eqn}
\end{equation}
where $\alpha$ and $\beta$ are functions of $r$ and $\theta$,
and ${\VEC {\nabla}}\times {\VEC {a}}={\VEC {\nabla}}\alpha \times{\VEC {\nabla}}\beta\ne0$ is assumed.
Owing to almost arbitrary functions $\alpha$ and $\beta$,
the constraint on the electric current, and hence to the 
magnetic-field configuration, is relaxed in the non-barotropic case.
Non-barotropicity has been studied in magnetic deformation
\citep{2011MNRAS.417.2288M,2013MNRAS.434.1658M,2015MNRAS.447.3475M}.
Barotropic 
\citep[][]{2008MNRAS.385..531H,2021MNRAS.502.2097K}
and non-barotropic models are significantly different.
\citep{2011MNRAS.417.2288M,2013MNRAS.434.1658M,2015MNRAS.447.3475M}
up to approximately
one order of magnitude in the resulting ellipticity.
The effect is important; however, the treatment remains unclear.
Therefore, we introduce the models of ${\VEC {\nabla}}\times {\VEC {a}}$
to study non-barotropicity in Section 4.

   \subsection{Magnetic-field Evolution}
 Our consideration is limited to the inner crust of a neutron star,
where the mass density ranges from 
$\rho_{c} =1.4\times 10^{14}$ g cm$^{-3}$ at the core--crust boundary $r_{c}$ 
to the neutron-drip density $\rho_{1} = 4\times 10^{11}$ g cm$^{-3}$ at $R=12$ km.
We ignore the outer crust compared to ``ocean''
and treat the exterior region of $r> R$ as the vacuum.
The crust thickness $\Delta r \equiv R-r_{c}$ is assumed to be $\Delta r/R=0.05$, $\Delta r=0.6$ km.
The Lorentz force ${\VEC {j}}\times {\VEC {B}}$ 
due to the magnetic-field evolution is not fixed in a secular timescale.
The evolution in the crust is governed by the induction equation
\begin{align}
\frac{\partial}{\partial t}{\VEC {B}}=-{\VEC {\nabla}}\times
\left[\frac{1}{e n_{\rm e}}{\VEC {j}}\times {\VEC {B}}+
\frac{c}{\sigma}{\VEC {j}}\right],
  \label{Frad.eqn}
\end{align}
where $n_{\rm e}$ is the electron-number density, and $\sigma$ is the electric conductivity.
The first term in Equation~(\ref{Frad.eqn}) represents the Hall drift, and
the Hall timescale $\tau_{\rm H}$ is estimated as follows:
\begin{align}
 \tau_{\rm H} =\frac{4\pi e n_{\rm e c} (\Delta r)^2}{c B_{0}}
=7.9 \times 10^5~{\rm yr}
\left(\frac{ B_{0} }{10^{14}{\rm G} }\right)^{-1} , 
\label{timescale.hall}
\end{align}
where the electron-number density 
$n_{\rm ec}$ at the core--crust boundary, 
crust thickness $\Delta r$, and dipole field strength $B_{0}$ at the surface are used.
This timescale is shorter than that of the Ohmic decay,
which is the second term in Equation~(\ref{Frad.eqn}) in the strong-magnetic-field regime.
The Hall-Ohmic evolution was numerically simulated in the Hall timescale 
\cite[][]{2007A&A...470..303P,2012MNRAS.421.2722K,2013MNRAS.434..123V,
2013MNRAS.434.2480G,2014MNRAS.438.1618G,
2021CoPhC.26508001V}
for axially symmetric models. Recently, the calculation has been extended to 3D models
\cite[][]{2015PhRvL.114s1101W,2019CoPhC.237..168V,2020ApJ...903...40D,
2020MNRAS.495.1692G,2021ApJ...909..101I},
revealing some of the effects ignored in the 2D models.
Here, our calculation is limited to the early phase of the evolution in a simpler axial-symmetric model.
We consider only the Hall drift term in Equation~(\ref{Frad.eqn}) and rewrite the equation as follows:
\begin{equation}
\frac{\partial}{\partial t}{\VEC {B}}=-{\VEC {\nabla}}\times
\left[
\frac{1}{e n_{\rm e}}{\VEC {j}}\times {\VEC {B}}
\right]
= -{\VEC {\nabla}} \chi \times {\VEC {a}}
 - \chi  {\VEC {\nabla}}\times{\VEC {a}},
%
 \label{BHevl.eqn}
\end{equation}
where
\begin{equation}
\chi \equiv \frac{c\rho}{e n_{e}}
= \frac{4\pi \rho_{c} (\Delta r)^2}{\tau_{\rm H}B_{0}}{\hat \chi}.
\label{chidef.eqn}
\end{equation}

In Equation~(\ref{chidef.eqn}),
${\hat \chi}$ is a dimensionless function,
which represents an inverse of the electron fraction.
 The electron-number density is obtained
 from the proton fraction of the equilibrium nucleus in 
 "cold catalyzed matter," i.e., it is determined in the ground state at $T=0$ K.
The data given by \cite{2001A&A...380..151D}
is approximately fitted by a smooth function: 
\begin{equation}
{\hat \chi} =\frac{ {\hat \rho}^{1/2}}{0.32+0.66  {\hat \rho}},
 \label{chiftfm.eqn}
\end{equation}
where ${\hat \rho}\equiv\rho/\rho_{c}.$
The spatial-density profile of a neutron star in $r_c\le r \le R$ is approximated as follows~\citep{2019MNRAS.486.4130L}:
\begin{equation}
{\hat \rho}=\frac{\rho}{\rho_{c}}
=\left[1-\left(1-\left(\frac{\rho_1}{\rho_{c}}\right)^{1/2}
 \right)\left(\frac{r-r_{c}}{\Delta r} \right)\right]^2 .
 \end{equation}
The radial derivative $d{\hat \chi}/dr=(d{\hat \chi}/d\rho)(d\rho/dr)$ sharply changes,
owing to an abrupt decrease in the density near the surface; however,
${\hat \chi}$ is a smoothly varying function of $\mathcal{O}(1)$.
This functional behavior, which originates from the stellar-density 
profile that is inherent in neutron stars, is crucial
in our numerical calculation.
Different fitting formulae are discussed for different equation of state in
\cite{2018MNRAS.481.2994P}; however,
the difference in ${\hat \chi}$ is not significant in our analysis.
We consider the early phase of the evolution 
in the axially symmetric equilibrium model, in which $a_{\phi}=0$.
From  Equation~(\ref{BHevl.eqn}) at $t=0$, we obtain that
${\partial B_{\phi}}/{\partial t} \neq 0$, but 
${\partial B_{r}}/{\partial t}={\partial B_{\theta}}/{\partial t}=0$.
The azimuthal component $B_{\phi}$ changes linearly with time $t$, whereas
the poloidal components change with $t^2$.
We limit our consideration to the lowest order of $t$ only
  and ignore the change in the poloidal magnetic field. 
  The early phase of the toroidal magnetic field may be approximated as
\begin{equation}
 \delta B_{\phi}=\frac{\delta S}{\varpi}\left(\frac{t}{\tau_{\rm H}}\right),
  \label{delB31.eqn}
\end{equation}
where $\delta S$ is a function of $r$ and $\theta$. 
Because it is associated with $\delta B_{\phi}$, the poloidal current changes; thus, the Lorentz force
$\delta {\VEC {f}}=c^{-1}(\delta{\VEC {j}} \times {\VEC {B}}+
{\VEC {j}} \times \delta{\VEC {B}})$  also changes.
We observe that the non-zero component is $\delta f_{\phi}$ because 
${\VEC{j}}_{p}=B_{\phi}=0$ at $t=0$, and we explicitly write 
\begin{equation}
    \delta f_{\phi} = \frac{1}{4\pi \varpi^2}
    [ {\VEC {\nabla}}\delta S \times  {\VEC {\nabla}} \Psi]_{\phi}
    \left(\frac{t}{\tau_{\rm H}}\right).
    \label{ScrossG.eqn}
\end{equation}
%

   \subsection{Quasi-stationary Elastic Response}
We assume that the solid crust elastically acts 
against the force $\delta f_{\phi}$.
The change is so slow that the elastic evolution is quasi-stationary. 
The acceleration $\partial^2 \xi_{i}/\partial t^2$ 
 of the elastic displacement vector $\xi_{i}$ is dismissed; thus,
the elastic force is balanced with the change in the Lorentz force, i.e., when
the gravity and pressure in Equation~(\ref{Forcebalance.eqn}) is assumed to be fixed.
The elastic force is expressed by the trace-free strain tensor $\sigma_{ij}$ and a shear modulus $\mu$;
therefore, the force balance is 
\begin{align}
{\nabla}_{i} \left(2\mu \sigma^{i}_{\phi} \right)
+ \delta f_{\phi}=0,
\label{ElasticForce.eqn}
\\
 \sigma_{ij} =\frac{1}{2}({\nabla}_{i} \xi _{j}  +{\nabla}_{j} \xi _{i}),
\end{align}
where incompressible motion ${\nabla}_{k} \xi ^{k}=0 $ is assumed. Alternatively, the equivalent form is
\begin{equation}
\nabla_{i} \left[2\mu \sigma^{i\phi} + \frac{1}{4\pi} B^{i} \delta B^{\phi} \right]
= 0.  
\label{ConsElMag.eqn}
\end{equation}

The relevant component induced by $\delta f_{\phi}$ is the azimuthal displacement $\xi_{\phi}$ only,
and the shear tensors that are determined by it are 
\begin{align} 
\sigma_{r \phi} =\frac{r}{2}\left(\frac{\xi_{\phi}}{r}\right)_{,r},
~~~~
\sigma_{\theta \phi} =\frac{\sin\theta}{2r}
\left(\frac{\xi_{\phi}}{\sin\theta}\right)_{,\theta} .
  \label{sigmax2.eqn}
\end{align}
The shear modulus $\mu$ increases with density, and it may be
approximated as a linear function of $\rho$,
which is overall fitted to the results of a detailed calculation reported in a previous study
\citep[see  Figure~43 in ][]{2008LRR....11...10C}.
 \begin{equation}
    {\mu}=\frac{\mu_{c}\rho}{\rho_{c}} ,
  \label{DFshMD.eqn} 
 \end{equation}
where $\mu_{c}=10^{30}~{\rm erg~cm}^{-3}$ at the core--crust interface.
The shear speed $v_{s}$ in Equation~(\ref{DFshMD.eqn}) 
is constant through the crust:
 \begin{equation}
   v_{s} = \left(\frac{\mu}{\rho} \right)^{1/2}=
   8.5\times 10^{7} {\rm {cm~s}^{-1}}.
\label{DFshVL.eqn} 
 \end{equation}
To solve Equation~(\ref{ElasticForce.eqn}),
 we use an expansion method with the Legendre polynomials
$P_l(\cos\theta)$ and radial functions $k_l(r)$, $a_{l}(r)$ as follows:
\begin{align}
   & \xi_{\phi}= -\sum_{l\ge 1} r k_{l} P_{l,\theta}
\left(\frac{t}{\tau_{\rm H}}\right),
\label{expndx3.eqn}
\\
& \delta f_{\phi}= -\sum_{l\ge 1}  r^{-3} a_{l} P_{l,\theta}
  \left(\frac{t}{\tau_{\rm H}}\right).  
 \label{expndf3.eqn}
\end{align}
The displacement $\xi_{\phi}$ is decoupled with respect to the index $l$,
owing to spherical symmetry, i.e., $\mu(r)$. 
Equation (\ref{ElasticForce.eqn}) is reduced to a set of ordinary differential equations for $k_{l}$
\citep{2022MNRAS.511..480K}:
\begin{equation}
 (\mu r^{4} k_{l}^{\prime})^{\prime} -(l-1)(l+2)\mu r^{2} k_{l}
    =-a_{l},
 \label{elaskl.eqn}    
\end{equation}
where a prime $^\prime$ denotes a derivative with respect to $r$.
The boundary conditions for the radial functions $k_{l}$ are given by the force-balance across the surfaces at $r_c$ and $R$.
That is, the shear-stress tensor $\sigma_{r\phi}$ vanishes because
other stresses for the fluid and magnetic field are assumed to be continuous.
Explicitly, we have 
$k_{l}^{\prime}=0$ both at $r_c$ and $R$.
Note that a mode with $l=1$ is special in Equation~(\ref{elaskl.eqn}), and 
$k_{1}$ is simply obtained by integrating $a_{1}$ with respect to $r$.

\section{Elastic Deformation in Barotropic Model}
The magnetic-field evolution in Equation~(\ref{BHevl.eqn}) is driven by two terms, which are separately examined.
We first consider the barotropic case, in which $ {\VEC \nabla}\times{\VEC a} =0$.
The evolution is driven by the first term in Equation~(\ref{BHevl.eqn}), 
i.e., the distribution of the electron fraction.
The linear growth term 
$\delta S$ in Equation~(\ref{delB31.eqn}) is obtained using  Equation~(\ref{EMForcebaro.eqn}) as
\begin{equation}
\delta S=\frac{2K_{0}\rho_{c}({\Delta}r)^2}{3B_{0}} 
{\hat{\chi}}^{\prime} g \sin \theta P_{2,\theta}.
   \label{barsfn.eqn} 
\end{equation}
In the case where the poloidal magnetic field ($\Psi=g\sin^2 \theta$)
is confined in the crust,
the constant $K_{0}$ is numerically obtained as
$K_{0}= 6.0 \times B_{0}/(\rho_{c} ({\Delta}r)^2)$.
Alternatively, we express  
$K_{0}=8.6 \times 10^1 v_{a}^2/(B_{0}R^2)$, where
$v_{a}$ is the {Alfv{\'e}n} speed in terms of $B_{0}$ and $\rho_{1}$:
\begin{equation}
    v_{a} =\frac{B_{0}}{\sqrt{4\pi\rho_1}}
    =4.5 \times 10^{7}  \left(\frac{B_{0} }{10^{14}{\rm G}} \right)
    {\rm {cm~s}^{-1}}.
  \label{DFalfVL.eqn} 
\end{equation}
The Lorentz force $\delta f_{\phi}$ in Equation~(\ref{ScrossG.eqn}) 
is calculated using Equation~(\ref{barsfn.eqn}).
For later convenience, we consider a general form
$\delta S=y_{l}(r)\sin \theta P_{l,\theta}$.
By using an identity for the Legendre polynomials, we reduce
Equation~(\ref{ScrossG.eqn}) to
\begin{align}
    \delta f_{\phi}=&\frac{1}{4\pi r^3} 
 \bigg[ (lg^\prime y_{l} +2g y_{l}^\prime)
\frac{l+1}{2l+1} P_{l-1,\theta}
+(- (l+1)g^\prime y_{l} +2g y_{l}^\prime )
\frac{l}{2l+1} P_{l+1,\theta} \bigg]\left(\frac{t}{\tau_{\rm H}}\right). 
 \label{flislpm1.eqn}   
\end{align}
Thus, the radial functions $a_{l-1}$ and  $a_{l+1}$
in Equation~(\ref{expndf3.eqn}) are induced by $y_{l}$.
By numerically solving Equation~(\ref{elaskl.eqn}) for $k_{l}$'s,
we obtain $\xi_{\phi}$ in Equation~(\ref{expndx3.eqn}) and the shear stress tensors,
$\sigma_{r\phi}$ and $\sigma_{\theta \phi}$, in  Equation~(\ref{sigmax2.eqn}).
For Equation~(\ref{barsfn.eqn}), the results are expressed using a combination of $k_{1}$ and $k_{3}$. 

\subsection{Results}
The shear stress increases homologously with time, 
i.e., the spatial profile of the shear force is
unchanged, but the magnitude increases with time.
The numerical calculation provides the maximum shear stress $\sigma_{\rm{max}}$ 
with respect to $(r,\theta)$ in the crust as follows:
\begin{equation}
\sigma_{\rm{max}}=6.5 \times 10^{1}\left(\frac{v_{a}}{v_{s}}\right)^2
\left(\frac{t}{\tau_{\rm{H}}}\right).
\end{equation}
The maximum is determined using
a ratio of the shear speed $v_{s}$ in Equation~(\ref{DFshVL.eqn})
to the {Alfv{\'e}n} speed $v_{a}$ in  Equation~(\ref{DFalfVL.eqn}).
Elastic equilibrium is possible, 
only when the shear strain satisfies a particular criterion.
We adopt the following (the Mises criterion)
to determine the elastic limit:
\begin{equation}
\frac{1}{2}\sigma_{ij}\sigma^{ij}
\le (\sigma_{\rm{max}})^{2}\le (\sigma_{c})^{2},
   \label{criterion}
\end{equation}
where $\sigma_{c}$ is a number  
$\sigma_{c} \approx 10^{-2}-10^{-1}$
\citep{2009PhRvL.102s1102H,
 2018PhRvL.121m2701C,2018MNRAS.480.5511B}. 
Thus, the period of the elastic response is limited by the constraint $\sigma_{c}$: 
\begin{align}
 \label{limitVA.eqn}
    t \le t_{*} & \equiv 
 1.5 \times 10^{-3}
  \left(\frac{ \sigma_{c}}{0.1}\right)\left(\frac{v_{s}}{v_{a}} \right)^2
\tau_{{\rm H}},
\\
&=4.3 \times 10^{3}
\left(\frac{ \sigma_{c}}{0.1}\right)
 \left(\frac{B_{0} }{10^{14}{\rm G}}  \right)^{-3}
{\rm{yr}}.
\end{align}
The breakup time $t_{*}$ becomes short, i.e., a few years for magnetars with
 $B_{0} =10^{15}$G.
 This is in agreement with the recurrent time of the activity in magnetars.
However, the timescale exceeds 1 Myr for most neutron stars
with $B <10^{13}{\rm G}$.
Moreover, other evolution effects are important, 
and present results are not applicable.

\subsection{Energy}
The stored elastic energy is obtained by
numerically integrating over the entire crust:
\begin{align}
 \Delta E_{\rm ela} &=
2\pi \int_{r_{c}} ^{R} r^2dr \int_{0} ^{\pi} \sin \theta d\theta
~\mu \sigma_{ij} \sigma^{ij}  
\nonumber
\\
&=5.8 \times 10^{-7}~\mu_{c}R^3
 \left(\frac{\sigma_{c}}{0.1}\right)^2
 \left( \frac{t}{t_{*}}\right)^2,
    \label{Elsenergy.eqn}
\end{align}
where we have used $R^3$ instead of $R^2{\Delta{r}}$
to normalize the crustal volume because ${\Delta{r}}/R=0.05$ is fixed.
The elastic energy $E_{\rm ela}$ increases up to $\approx 10^{41}$~erg 
at the breakup time $t_{*}$.
The magnetic energy $\Delta E_{\rm mag}$ is also obtained by
\begin{align} 
\Delta E_{\rm mag} &=
2\pi \int_{r_{c}} ^{R} r^2dr \int_{0} ^{\pi} \sin \theta d\theta
~\frac{(\delta B_{\phi})^2 }{8\pi} 
\nonumber
\\
&=2.0 \times 10^{-4}~B_{0}^2 R^3
 \left(\frac{\sigma_{c}}{0.1}\right)^2\left(\frac{v_{s}}{v_{a}}\right)^4
 \left( \frac{t}{t_{*}}\right)^2.
 \label{Magenergy.eqn} 
\end{align}
Here, the shear speed appears in $\Delta E_{\rm mag}$ because the
breakup time $t_{*}$ in Equation~(\ref{limitVA.eqn}) is used instead of $\tau_{\rm H}$.
The magnetic energy $\Delta E_{\rm mag}$ at $t_{*}$ is 
$\Delta E_{\rm mag} \approx 2\times 10^{43} (B_{0}/10^{14}{\rm G})^{-2}$ erg. However, it is considerably smaller than
the poloidal magnetic energy $E_{\rm mag,p}$, which
is numerically calculated as 
$E_{\rm mag,p}=3.8 B_{0}^2 R^3\approx 4\times 10^{46} (B_{0}/10^{14}{\rm G})^{2}$~erg.
Note that total magnetic energy is conserved by the Hall evolution. 
Therefore, the same amount of polar magnetic energy decreases.
However, we ignored the change in the poloidal component
and its energy, which are proportional to $t^2$.
The ratio of Equations~(\ref{Elsenergy.eqn}) and (\ref{Magenergy.eqn}) is 
\begin{equation}
\frac{d{\Delta}E_{\rm ela}/dt}{d{\Delta}E_{\rm mag}/dt}
=\frac{\Delta E_{\rm ela}}{\Delta E_{\rm mag} }
=8.1 \times 10^{-2}
 \left(\frac{v_{a}}{v_{s}}\right)^2
 =2.3 \times 10^{-2} \left(\frac{B_{0} }{10^{14}{\rm G}}\right)^2,
\end{equation}
where $\mu_{c}$ and $B_{0}^2$ are eliminated using $v_{s}$ and $v_{a}$.
 The ratio is proportional to $B_{0}^2$; thus, 
 $\Delta E_{\rm mag}$ decreases 
 in more strongly magnetized neutron stars.
From the viewpoint of the energy flow from the poloidal component,
the breakup energy $\Delta E_{\rm ela} \approx 10^{41}$ at the terminal 
is fixed, but the $\Delta E_{\rm mag}$ stored in the middle
depends on the Hall drift speed.
 The elastic energy is efficiently accumulated through toroidal magnetic energy with an increase in $B_{0}$;
 $\Delta E_{\rm ela} >\Delta E_{\rm mag}$
 for $B_{0}> 6.7 \times 10^{14}$~G.

\subsection{Spatial Distribution}

\begin{figure}\begin{center}
  \includegraphics[width=0.5\columnwidth]{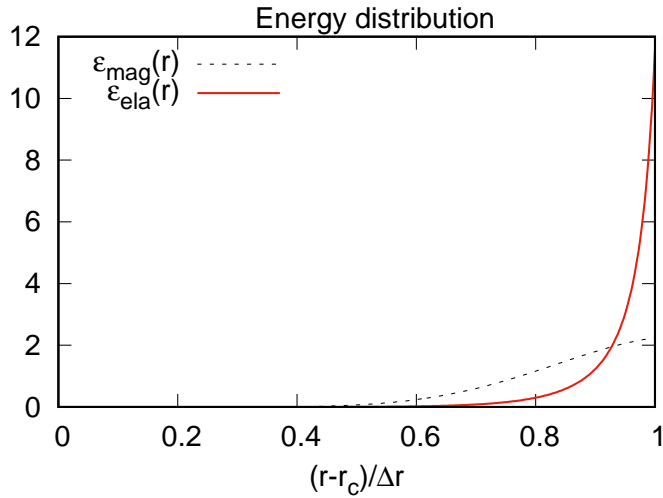}
\caption{ 
 \label{Fig1}
Energy distribution in crust as a function of the radius.
Normalized energy density
$\varepsilon (r)$ is plotted for 
 magnetic energy, which is denoted by a dotted curve, and elastic energy, denoted by a solid curve.
}
\end{center}\end{figure}

Figure~\ref{Fig1} shows spatial-energy densities
$\varepsilon_{\rm ela}(r)$ and $\varepsilon_{\rm mag}(r)$, with regard to 
$\Delta E_{\rm ela}$ and $\Delta E_{\rm mag}$
in the crust, respectively.
They are normalized as 
$\int \varepsilon_{\rm ela} dr=\int \varepsilon_{\rm ela} dr=1$.
Evidently, both energies are highly concentrated near the surface $r\approx R$.
This property comes from the radial derivative of ${\hat \chi}$ 
in Equation~(\ref{chidef.eqn}).
$d{\hat \chi}/dr=(d{\hat \chi}/d\rho)(d\rho/dr)$
is steep there even though ${\hat \chi}$ is $\mathcal{O}(1)$.
The large value comes from $|d\rho/dr|$, i.e.,
a sharp decrease in density near the stellar surface,
and it results in a smaller evolution timescale
$\ll \tau_{\rm{H}}$ in Equation~(\ref{limitVA.eqn}).
%

\begin{figure}\begin{center}
  \includegraphics[width=0.5\columnwidth]{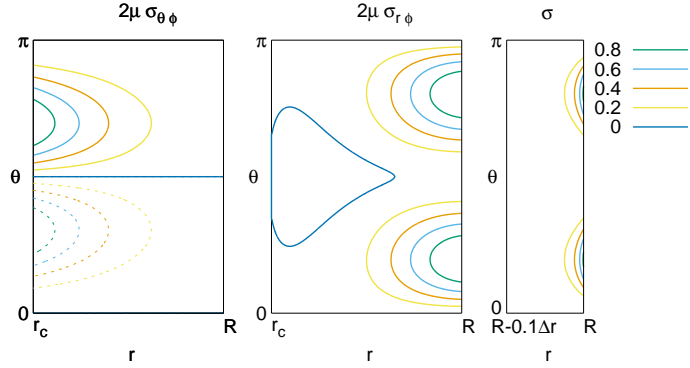}%
\caption{ 
 \label{Fig2}
Contour map of stress tensor in crust.
The left panel shows $2\mu \sigma_{\theta \phi}$,
the middle $2\mu \sigma_{r \phi}$, and
the right the magnitude $\sigma$.
They are normalized according to the maximum.
Negative values are plotted using a dotted curve in the left panel.
The magnitude $\sigma$ in the right panel
is shown in the outer small region near the surface,
$R-0.1{\Delta}r \le r \le R$.
}
\end{center}\end{figure}

Shear stresses $\sigma_{\theta \phi}$ and $\sigma_{r\phi}$
are induced by the axial displacement $\xi_{\phi}$.
A numerical calculation shows that
the component $\sigma_{r\phi}$ is considerably larger than 
 $\sigma_{\theta \phi}$; $(\sigma_{r\phi})_{\rm max}\sim 200 (\sigma_{\theta \phi})_{\rm max}$.
 Figure~\ref{Fig2} shows their spatial distribution using a contour map 
 of $2\mu \sigma_{\theta \phi}$ and $2\mu \sigma_{r\phi}$ in the $r-\theta$ plain.
The angular dependence of 
 $\sigma_{\theta\phi}$ is $\sigma_{\theta \phi} \propto \sin^2 \theta \cos \theta$, and it is
 anti-symmetric with respect to the equator ($\theta =\pi/2$).
Moreover,
$\sigma_{r\phi}$ is the sum 
of $P_{1,\theta}$ and  $P_{3,\theta}$, and it is symmetric
with respect to $\theta=\pi/2$.
The magnitude 
$\sigma = (\sigma_{ij} \sigma^{ij}/2)^{1/2}$
is also shown in the right panel, and $\sigma$
is sharp near the surface, as expected from
the sharp energy-density-distribution in Figure~\ref{Fig1}
We discuss the modification of the poloidal magnetic field at the core--crust boundary.
Thus far, the magnetic field is expelled there.
When the field is penetrated to the core, the function $g$ near the boundary and the constant $K_{0}$ in Equation~(\ref{barsfn.eqn}) are changed.
The former is unimportant because the function ${\hat{\chi}}^{\prime}$ is sharp near the surface, and this fact determines the result, as shown in Figures~\ref{Fig1} and \ref{Fig2}.
The constant $K_{0}$ for the penetrated field is 
$4.1 \times 10^{-2}$ times smaller than that for the expelled one.
Consequently, the profile is almost unchanged, but the break time $t_{*}$ increases by a factor of 24 with the same dipole field strength.
%

\section{Elastic Deformation in Non-Barotropic Model}
We consider the evolution driven by the second term, ${\VEC {\nabla}}\times {\VEC {a}}\ne 0$
in  Equation~(\ref{BHevl.eqn}), which originates from the non-barotropic material distribution.
However, ${\VEC {\nabla}}\times{\VEC {a}}$ and
the corresponding magnetic field cannot be easily estimated, unless the non-barotropic property is specified.
A large freedom of choice hinders our analysis.
Therefore, we simply model the term ${\VEC {\nabla}}\times {\VEC {a}}$ and
understand the non-barotropic effect in its magnitude and property.
For this purpose, we assume $a_{\phi}=0$ and 
 \begin{equation}%
  ({\VEC {\nabla}}\times {\VEC {a}})_{\phi}= \frac{N^2}{(\Delta r)^2}
  F_{n}(r)  P_{2, \theta},
 \label{nonbfmp2.eqn}
\end{equation}
 where $N$ is a constant that characterizes the non-barotropic strength, and it has
 the dimension of velocity. Additionally,
 $F_{n}$ is a non-dimensional radial function.
 We consider a small deviation from the barotropic case, for which
 the second term in Equation~(\ref{EMForceNB.eqn}) is smaller than the first term.
 Therefore, the magnetic field is approximated using the barotropic case,
 i.e., the poloidal magnetic function $\Psi$ and $S=0$.
 This treatment constrains
 the normalization $N$ in Equation~(\ref{nonbfmp2.eqn}) with respect to the magnitude.
 By the dimensional argument, we have 
 $|N| \ll R/t_{\rm{ff}}$ $\sim 10^{9}~{\rm{cm~s}}^{-1}$,
where $t_{\rm{ff}}$ is a free-fall timescale.
Moreover, $|N|<v_{a}$ and 
 $|N|<v_{s}$ $\sim 10^{8}~{\rm{cm~s}}^{-1}$ are also likely because
the crust is in magneto-elastic equilibrium.
 The angular dependence of Equation~(\ref{nonbfmp2.eqn})
is chosen for $\delta S$ to be the same as in Equation~(\ref{barsfn.eqn}).
 The radial function $F_{n}$ has a maximum that is normalized as unity, and it
 vanishes at inner and outer boundaries; the function is modeled as follows: 
\begin{equation}
F_{n} = \frac{256}{27}\left(\frac{r-r_{c}}{\Delta r} \right)^{n}
 \left( \frac{R-r}{\Delta r}  \right)^{4-n},
 \label{modelFN.eqn}
\end{equation}
where $n=1$ or 3.
The model with $n=1$ is referred to as the "in" model because the 
maximum is located at $r=r_{c}+\Delta r/4$, whereas
that with $n=3$ is referred to as the "out" model because the 
maximum is located at $r=R-\Delta r/4$.

   \subsection{Results of a Simple Model}
We neglect the first term in Equation~(\ref{BHevl.eqn}), and 
consider the magnetic-field evolution driven by 
the term ${\VEC {\nabla}}\times {\VEC {a}}$ (Equation~(\ref{nonbfmp2.eqn})) only.
Similar to the calculations in Section 3, 
a linearly growing shear--stress is obtained, owing to $\xi_{\phi}$.
The period of the elastic response is limited by
\begin{equation}
t \le t_{*} \equiv 
 n_{1} \times
  \left(\frac{ \sigma_{c}}{0.1}\right)\left(\frac{v_{s}}{N}\right)^2
\tau_{{\rm H}},
  \label{limitTN.eqn}    
\end{equation}
where $n_{1}$ is a number of the order of $10^{-2}$, depending on the models,
as listed in Table~\ref{table1:mylabel}.
Owing to our simple modeling,
the comparison between the barotropic and non-barotropic models is uncomplicated; 
the {Alfv{\'e}n} speed $v_{a}$ in Equation~(\ref{limitVA.eqn})
is formally substituted by $N$ in Equation~(\ref{limitTN.eqn}). 
The elastic energy $\Delta E_{\rm ela}$ 
and toroidal magnetic energy $\Delta E_{\rm mag}$ 
stored inside the crust are also summarized as follows:
\begin{align}
 &\Delta E_{\rm ela} =
n_{2} \times \mu_{c}R^3\left(\frac{\sigma_{c}}{0.1}\right)^2
 \left( \frac{t}{t_{*}}\right)^2,
  \label{Elsenergy2.eqn}
\\
&\Delta E_{\rm mag}
=n_{3} \times B_{0}^2 R^3
 \left(\frac{\sigma_{c}}{0.1}\right)^2
 \left(\frac{v_{s}}{N}\right)^4\left( \frac{t}{t_{*}}\right)^2,
    \label{Magenergy2.eqn}
\end{align}
where $n_{2}\approx  6 \times 10^{-4}$, and
$n_{3}\approx  10^{-6}$, as listed in Table~\ref{table1:mylabel}.
The elastic energy $\Delta E_{\rm ela}$ 
does not depend on $N$; however, the timescale (\ref{limitTN.eqn}) does.
The amount of elastic energy is unrelated to the detailed process, which affects the accumulation speed in the crust.
%
At the breakup time, the elastic energy is 
$\Delta E_{\rm ela} =2\times 10^{44} -2\times 10^{45}$ erg.
This energy is more than three orders of magnitude
larger than the energy (\ref{Elsenergy.eqn}) considered in the previous section.
The difference is made clear when considering the energy--density distribution.

\begin{table}
    \centering
    \begin{tabular}{lccc}
model &  $n_1~~(t_{*})$ &   $n_2~~(\Delta E_{\rm ela})$ & $n_3~~(\Delta E_{\rm mag})$\\
\hline \hline
 in & $8.0 \times 10^{-3}$ &
 $2.1 \times 10^{-3}$&
 $7.2\times 10^{-7}$
   \\
  out &
    $1.2 \times 10^{-2}$&
    $2.4 \times 10^{-4}$&
    $1.0 \times 10^{-6}$
    \\ \hline
ave &
    $3.6 \times 10^{-3}$&
    $2.9 \times 10^{-4}$&
    $2.9 \times 10^{-7}$
    \\
min &
    $1.9 \times 10^{-3}$&
    $0.49 \times 10^{-4}$&
    $0.73 \times 10^{-7}$
    \\
max &
    $5.5 \times 10^{-3}$&
    $7.0 \times 10^{-4}$&
    $6.4 \times 10^{-7}$
    \\
    \hline \hline
    \end{tabular}
    \caption{Numerical values in eqs.~(\ref{limitTN.eqn}),
    (\ref{Elsenergy2.eqn}), and
    (\ref{Magenergy2.eqn}). }
\label{table1:mylabel}
\end{table}

\begin{figure}\begin{center}
  \includegraphics[width=0.5\columnwidth]{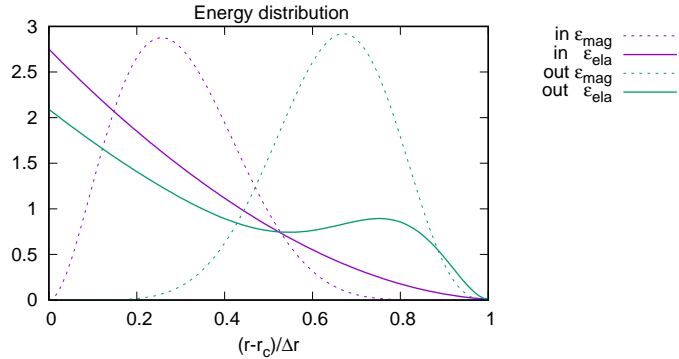}
\caption{ 
\label{Fig3}
Spatial distribution of magnetic energy (dotted curve) and
elastic energy (solid curve) in the crust.
Normalized energy densities $\varepsilon (r)$ are plotted for two models.
}
\end{center}\end{figure}

Figure \ref{Fig3} shows the energy--density distribution in the crust.
The difference in the toroidal magnetic energy clearly originates 
in the model choice;
the energy density spreads more towards the interior for the "in" model, whereas it spreads 
more towards the exterior for the "out" model.
Most of the elastic energy is localized near the inner core--crust boundary; however,
the distribution in the "out" model is shifted outwardly
with the second peak ($\sim r_c + 0.8{\Delta}r$) produced by the input model.
The integrated elastic energy in the "out" model
becomes one order of magnitude smaller than that in the "in" model.
The amount of elastic energy at
the breakup clearly depends on the spatial distribution of the energy density because the shear modulus $\mu$ is a strongly decreasing function toward the surface.
The elastic limit of the entire crust is typically determined using a condition
 to the shear $\sigma_{ij}$ near the surface.
 The total elastic energy $\sim \int \mu \sigma_{ij} \sigma^{ij} d^3x$ 
 thereby decreases as $\sigma_{ij} \sigma^{ij}$ is localized towards the exterior.
The breakup elastic energy $\Delta E_{\rm ela} \sim 10^{41}$~erg at $t_{*}$ 
in the previous section is an extreme case because the energy density is concentrated near the surface.

\begin{figure}\begin{center}
 \includegraphics[width=0.5\columnwidth]{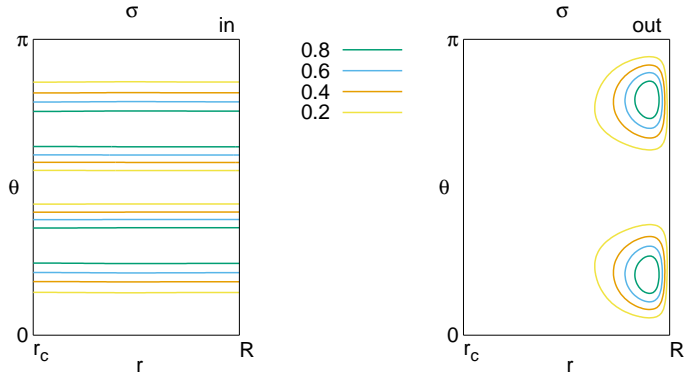}
\caption{ 
\label{Fig4}
Crust contour map of magnitude of stress tensor normalized using the maximum. 
The left panel for the "in" model shows $\sigma\approx\sigma_{\theta \phi}$, whereas
 the right panel for the "out" model shows $\sigma\approx\sigma_{r\phi}$.
}
\end{center}\end{figure}

Figure~\ref{Fig4} shows the 
magnitude of shear stress $\sigma$ inside the crust. 
The contours of $\sigma$ in the two models are different.
We identified that the dominant component in the "in" model (left panel) is 
$\sigma_{\theta \phi}$, which has an angular dependence that is described by
$\sigma_{\theta \phi} \propto \sin^2 \theta \cos \theta$.
 The maximum of $\sigma$ is given along a line
$\cos\theta= \pm 1/\sqrt{3}~(\theta \sim 55^{\circ}, 125^{\circ})$.
The component $\sigma_{r \phi}$ is dominant in the "out" model (right panel). 
Sharp peaks are localized near the surface, similar to
the right panel in  Figure~\ref{Fig2}; however,
the localization is not as pronounced as in  Figure~\ref{Fig2}.
The magnitude $\sigma$,
which is important to determine the critical limit, is large near the surface.

  \subsection{Results of a Model Including Higher Multipoles}

%
In a more realistic situation, the solenoidal acceleration may fluctuate spatially.
We consider the sum of multi-pole components $P_{l, \theta}$:
\begin{equation}%
  ({\VEC {\nabla}}\times {\VEC {a}})_{\phi}= \frac{N^2}{(\Delta r)^2}
  \sum_{l=2} ^{l_{\rm max}} \left(\frac{l}{2}\right)^{-4/3}
\lambda_{l} F_{n}(r)  P_{l, \theta},
  \label{nonbfmpsm.eqn}
\end{equation}
where 
$\lambda_{l} F_{n}$ is a radial function that is randomly selected from
 $\pm F_{1}$ or $\pm F_{3}$ depending on $l$.
 As discussed for Equation~(\ref{flislpm1.eqn}),
 the radial function $k_{l}$ in the azimuthal displacement $\xi_{\phi}$
 is solved for the source term
 that originates from $\lambda_{l-1} F_{n}$ and $\lambda_{l+1} F_{n}$; thus,
 the amplitude $|k_{l}|$ fluctuates according to the randomness.
We fix the overall constant $N$. 
Equation~(\ref{nonbfmpsm.eqn}) is reduced to Equation~(\ref{nonbfmp2.eqn})
when $l_{\rm max}=2$. Moreover, higher $l$-modes, up to $l_{\rm max}=30$
with the power-law weight, are included.
The power-law index is considerably steep; therefore, 
the dominant component is still described by $l=2$.
We calculated 20 models by randomly mixing $\lambda_{l} F_{n}$.
The numerical results are summarized in the same forms as
eqs.~(\ref{limitTN.eqn})--(\ref{Magenergy2.eqn}), and the
numerical values
$n_{i}~(i=1,2,3)$ in the breakup time and energies are listed in Table~\ref{table1:mylabel} according to 
the average, minimum, and maximum for the 20 models.
These numerical values are of the same order as
those by a single mode with $l=2$ because we include higher $l$-modes as the correction.
Interestingly, the breakup time $t_{*}$ generally becomes shorter than
that for a single mode with $l=2$ because the higher modes $l>3$ are cooperative.
%

\begin{figure}\begin{center}
   \includegraphics[width=0.7\columnwidth]{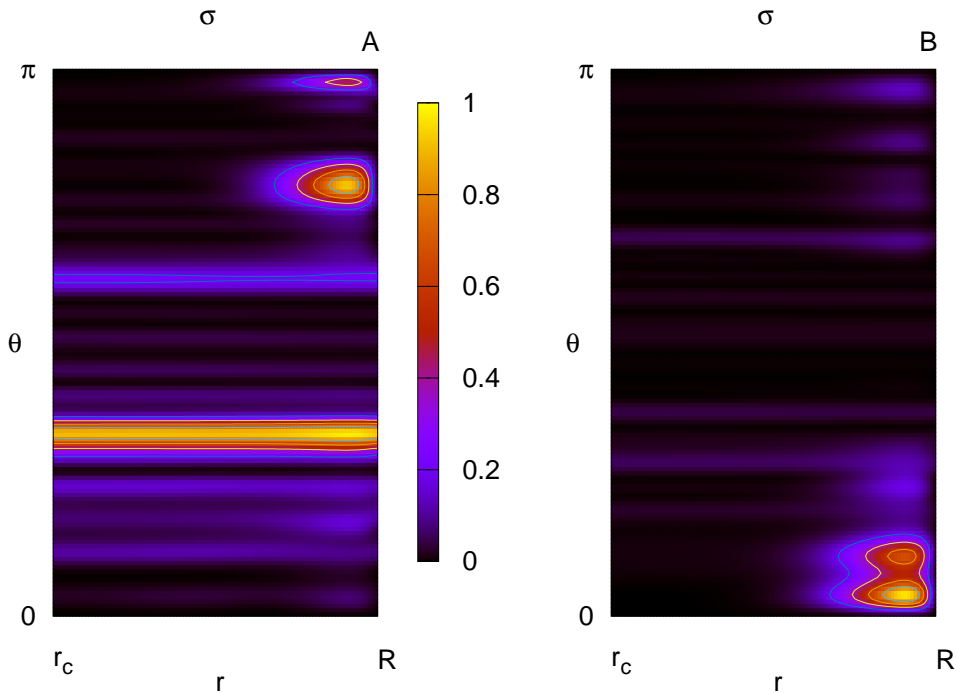}
\caption{
\label{Fig5}
Color contour map of crust for the magnitude of the stress tensor $\sigma$
normalized using the maximum; two models are compared. 
}
\end{center}\end{figure}

Figure~\ref{Fig5} demonstrates the spatial distribution of the shear stress tensor.
Two models are shown using contours of the magnitude of $\sigma$.
In the left panel, the sub-critical regions are on 
a constant $\theta$ line with a sharp peak near the surface.
In the other model (right panel), 
a peak is observed at $\theta =0$ near the surface.
The angular position of the peak is different between the two models.
As shown in  Figure~\ref{Fig4}, 
 the spatial pattern along a constant $\theta$ comes from the component $\sigma_{\theta \phi}$,
 whereas the sharp peak near the surface is due to $\sigma_{r \phi}$.
The mixing of the two types of radial functions, $F_{1}$ and $F_{3}$,
and angular functions $P_{l}$ with random phases and weights
only complicates the spatial distribution of $\sigma$.
A sharp peak is likely to be located near the surface.
The outer part of the crust is always fragile; thus, the breakup time becomes shorter.
%

 \section{Summary and Discussion}
We have considered the evolution of elastic deformation over a secular timescale
($>1$~yr) starting from zero displacement.
The initial state is related to the dynamic force balance
that is determined within a second.
When a neutron star cools below the melting temperature $T \sim 10^9$~K,
its crust is crystallized.
Meanwhile, the pressure, gravity, and Lorentz force 
are balanced without the elastic force.
In another situation, 
the elastic energy settles to the ground state, and zero displacement occurs
after the energy is completely released at crustal fracture.
Therefore, the initial condition is simple and natural.
%

When the MHD equilibrium is axisymmetric,
the azimuthal component of the magnetic field increases linearly according to the Hall evolution.
Consequently, the elastic deformation in the azimuthal direction
is induced to cancel the change in the Lorentz force, and
the shear strain gradually increases.
We estimate the range of the elastic response.
Beyond the critical limit, the crust responds plastically or fractures.
Our calculations provide the breakup time and shear distribution at the threshold.

For the barotropic case, the breakup time until fracture is proportional to the cube of the magnetic-field strength.
The time becomes a few years for a magnetar with a surface dipole of $B_{0} \sim 10^{15}$G, when the field is located outside the core.
However, it exceeds 1 Myr for most radio pulsars ($B_{0} <10^{13}$G), 
and the process is irrelevant to them.
In addition to the field strength, the timescale is typically shortened 
by a factor of $10^{-3}$ smaller than the Hall timescale because the elastic displacement is highly concentrated near the surface.
The driven mechanism is related to 
the instability associated with electron-density gradients \citep{2014PhPl...21e2110W}.
The distribution in any realistic model of neutron-star crusts is considerably sharp; therefore,
the evolution is general. 
Another type of Hall-drift instability occurs in the presence of a non-uniform 
magnetic field \citep{2002PhRvL..88j1103R}, which is not considered here,
and its energy would be smaller, owing to the size of the irregularity.
In our calculation, we do not follow the instability; instead, we estimate
the energy transferred to the elastic deformation.

The elastic energy at the critical limit 
in the model driven by the electron-number-density gradient is $\sim 10^{41}$ erg.
The amount of energy is of the same order as that of short bursts in magnetars.
The breaktime of $\sim 10$ years also reconciles with the observed recurrent-time of the bursts.
 However, the energy $\sim 10^{41}$ erg 
 is smaller than that of giant flares $\sim 10^{44} -10^{46}$ erg
\citep[][for a review]{2015RPPh...78k6901T,2017ARA&A..55..261K,2021ASSL..461...97E}.
The total elastic energy derived in Section 3 is based on the electron-number density in cold catalyzed matter,
i.e., the ground state at $T=0$ K.
If the assumption was relaxed, the non-barotropic effects might
increase the total elastic energy, as considered in Section 4.

When the pressure distribution is no longer expressed solely by 
the density $\rho$,
the magnetic evolution is affected by the solenoidal acceleration
${\VEC {a}}={\VEC {\nabla}} \times( \rho^{-1}{\VEC {\nabla}}P)\ne 0$.
We have also considered this effect 
by creating the model in terms of a spatial function and an overall strength parameter, which are assumed to be constant 
in time in our non-barotropic model.
Using the simplified model, we calculated the breakup time of the crustal failure and the
energies stored in the crust. The results were comparable to those for the barotropic case.
The strength parameter significantly affects the breakup time;
the larger the magnitude, the shorter the breakup time.
However, the amount of elastic energy at the breakup 
does not depend on the strength parameter, but only on the spatial function.
The maximum elastic energy considerably increases up to $\sim 10^{45}$ erg.
However, the model is still primitive, and thermal evolution should  
 also be incorporated to investigate a more realistic situation.

The maximum energy has been explored thus far; however, 
a natural question arises. What is the fraction of energy that is released at 
the crustal fracture when the strain exceeds the threshold?
This question is important, but at present unclear, owing to our lack of understanding of the fracture dynamics.
Therefore, we present the following discussion.
As depicted in Figure~\ref{Fig5}, in the realistic mixture model, 
a peak of shear strain $\sigma$ is probably located near the surface, 
where the crust is fragile.
Therefore, the fracture should not include the whole crust, but only the shallow crust.
For this case, the released energy is not 
the whole elastic energy $\sim 10^{45}$ erg, 
but the energy stored in the restricted region, i.e.,
a small fraction of the total, probably $\sim 10^{41}$ erg.
%

The elastic deformation driven by the Hall evolution is simulated for the first time.
The critical structure at the breakup time is crucial for subsequent evolution, 
irrespective of plastic evolution or fracturing.
The transition may appear as a burst on a magnetar.
The magnetic-field rearrangement due to a mimic burst was incorporated in the Hall evolution
\citep{2011ApJ...741..123P,2013MNRAS.434..123V,2020ApJ...902L..32D},
without solving elastic deformation.
These studies estimated the critical state based on the magnetic stress ${\mathcal{M}}_{ij}$.
In numerical simulations,
${\mathcal{M}}_{ij}$ changed, and the critical state was assumed
when a condition among ${\mathcal{M}}_{ij}$ reached the threshold value.  
Similar approximations for the elastic limit,
 which were derived solely from ${\mathcal{M}}_{ij}$,
were used in a previous study
 \citep[][]{2015MNRAS.449.2047L,2019MNRAS.486.4130L,2019MNRAS.488.5887S}.
Mathematically, the shear stress $\sigma_{ij}$ 
cannot be derived from ${\mathcal{M}}_{ij}$ without solving 
the appropriate differential equation
$\nabla_{j}(2\mu\sigma_{i} ^{j} +{\mathcal{M}}_{i}^{j})=0$
(see Equation~(\ref{ConsElMag.eqn})).
 Therefore, previous results with the criterion ${\mathcal{M}}_{ij}$ 
are questionable.

Our calculation shows that the period of elastic evolution is typically
 $10^{-3}$ times the Hall timescale; however, this value depends on
 the strength and geometry of the magnetic field.
 The timescale is shorter than the Ohmic timescale for $B\ge 10^{13}$ G.
The magnetic-field evolution beyond the period may be described by 
including the viscous bulk flow when the crust responds plastically.
The effect of the plastic flow on the Hall--Ohmic evolution was considered
by assuming a plastic flow everywhere in the crust \citep{2020MNRAS.494.3790K}
or by using an approximated criterion
\citep{2019MNRAS.486.4130L,2021MNRAS.506.3578G,2022arXiv220206662G}.
The effect may be regarded as additional energy lost to the Ohmic decay.
However, the post-failure evolution significantly depends on the modeling 
 in the numerical simulation \citep{2021MNRAS.506.3578G,2022arXiv220206662G}.
That is, the region of plastic flow is either local or global
when the failure criterion is satisfied.
Therefore, the manner of incorporation of crust failure in the numerical simulation must be explored.
Finally, further investigation is required before the elastic deformation toward the crust failure
can be considered a viable model.
The effect of magnetic field configuration should be considered because there are many degrees of freedom concerning it. 
Moreover, the outer boundary, i.e., inner--outer crust boundary or exterior magnetosphere is crucial as the crust becomes more fragile with increasing radius.
Meanwhile, the electric conductivity decreases and the Ohmic loss becomes more important.
By considering coupling to an exterior magnetosphere,
twisting of the magnetosphere as well as crustal motion
will be calculated in a secular timescale to match astrophysical observations,
e.g., to describe the pre-stage of outbursts, like SGR 1830-0645
\citep[][]{2022ApJ...924L..27Y}.

\section*{Acknowledgements}
 This work was supported by JSPS KAKENHI Grant Number JP17H06361,JP19K03850.

  \bibliography{kojima22July}
\bibliographystyle{aasjournal} 

  \end{document}